\documentclass{optica-article}
\journal{boe}
\articletype{Research Article}
\usepackage{graphicx}
\usepackage{setspace}
\usepackage{tocloft}
\usepackage{comment}
\usepackage{color,xspace}
\usepackage{multirow, booktabs, makecell}
\hypersetup{bookmarksnumbered}

\newcommand{\etal}{\textit{et al.\@}\xspace}

\newcommand{\um}{\(\muup\)m\xspace}
\newcommand{\yascom}[1]{\textcolor{red}{\xspace}}

\newcommand{\figurenames}{Figs.}

\graphicspath{{figs/}}

\begin{document}
\title{Synthesizing the degree of polarization uniformity from non-polarization-sensitive optical coherence tomography signals using a neural network}
\author{Shuichi Makita,\authormark{1*} Masahiro Miura,\authormark{2} Shinnosuke Azuma,\authormark{3} Toshihiro Mino,\authormark{3} and Yoshiaki Yasuno\authormark{1}}
\address{\authormark{1}Computational Optics Group, University of Tsukuba, 1--1--1 Tennodai, Tsukuba, Ibaraki 305--8573, Japan\\
\authormark{2}Department of Ophthalmology, Tokyo Medical University Ibaraki Medical Center, 3--20--1 Chuo, Ami, Ibaraki 300--0395, Japan\\
\authormark{3}Topcon Corporation, 75--1 Hasunumacho, Itabashi, Tokyo 174--8580, Japan}
\email{\authormark{*}shuichi.makita@cog-labs.org}
\homepage{http://optics.bk.tsukuba.ac.jp/COG/}

\begin{abstract}
	Degree of polarization uniformity (DOPU) imaging obtained by polarization-sensitive optical coherence tomography (PS-OCT) has the potential to provide biomarkers for retinal diseases.
	It highlights abnormalities in the retinal pigment epithelium that are not always clear in the OCT intensity images.
	However, a PS-OCT system is more complicated than conventional OCT\@.
	We present a neural-network-based approach to estimate the DOPU from standard OCT images.
	DOPU images were used to train a neural network to synthesize the DOPU from single-polarization-component OCT intensity images.
	DOPU images were then synthesized by the neural network, and the clinical findings from ground truth DOPU and synthesized DOPU were compared.
	There is a good agreement in the findings for RPE abnormalities: recall was 0.869 and precision was 0.920 for 20 cases with retinal diseases.
	In five cases of healthy volunteers, no abnormalities were found in either the synthesized or ground truth DOPU images.
	The proposed neural-network-based DOPU synthesis method demonstrates the potential of extending the features of retinal non-PS OCT\@.
\end{abstract}

\section{Introduction}

Optical coherence tomography (OCT)\cite{huang_optical_1991} visualizes cross-sectional images of the posterior eyes non-invasively and provides morphological structures of the retinal layers.
OCT is used for diagnosis and follow-up retinal treatment\cite{costa_retinal_2006}.
Recently, degree-of-polarization uniformity (DOPU) imaging~\cite{gotzinger_retinal_2008-1, baumann_polarization_2012, makita_degree_2014} using polarization-sensitive OCT (PS-OCT) has been used to investigate abnormalities of the retinal pigment epithelium (RPE)\cite{pircher_human_2006, michels_value_2008, baumann_segmentation_2010, ahlers_imaging_2010, schlanitz_performance_2011, lammer_imaging_2013, schutze_lesion_2013, ritter_characterization_2013, sayegh_polarization-sensitive_2015, schlanitz_identification_2015, augustin_multi-functional_2016,fialova_polarization_2016, roberts_retinal_2016, miura_evaluation_2017, miura_evaluation_2019, miura_evaluation_2019-2,augustin_optical_2020, miura_evaluation_2021}.
These studies have shown that DOPU imaging and DOPU analysis can be used for the detection and investigation of RPE anomalies.
Hence, the DOPU is a potential biomarker for RPE abnormalities.

Although the DOPU could improve retinal diagnosis, its widespread use is prevented by the extra cost of the PS-OCT setup.
The DOPU is a randomness metric of the polarization state of the backscattered probe beam.
To detect the polarization state of the backscattered probe light, two orthogonal polarization components should be detected.
This requires two detectors as well as a polarization splitter and controller in addition to a standard OCT setup.
In the case of swept-source OCT, the detection module can be simplified by multiplexing two polarization components~\cite{rivet_optical_2016,rivet_passive_2017,moon_fiber-based_2019,romodina_polarization_2022}.
For spectral-domain OCT, several methods have been proposed to use a single line-scan camera to acquire two polarization channels; parallel detection by multiplexing two different input angles to a spectrometer~\cite{baumann_single_2007}, by introducing a Wollaston prism~\cite{cense_polarization-sensitive_2007,mujat_autocalibration_2007,wang_polarization-maintaining_2010,wu_single_2020,cense_two_2022}, or by using a single multi-line scan camera~\cite{song_polarization-sensitive_2010,cense_two_2022}, time-division multiplexing~\cite{lee_high-speed_2010,park_buffered_2019}, and dual references for decoding two polarization components in optical path length~\cite{fan_spectral_2007} or in the spatial frequency domain~\cite{schmoll_single-camera_2010}.
However, it is inevitable to modify the optical setup and sacrifice some performance, such as imaging range or imaging speed.
If the DOPU could be obtained from a conventional OCT setup with a single polarization component detection, this would extend applications of the existing OCT devices.

Scattering of light by dense scatterers (multiple scattering)~\cite{bicout_depolarization_1994,morgan_effects_1997} and anisotropic scatterers~\cite{mishchenko_depolarization_1995} have been considered as possible causes of polarization randomization.
These scattering properties may affect the formation of speckles in OCT intensity images\cite{karamata_multiple_2005-1,hillman_detection_2010}.
However, it is not straightforward to determine their relationship, and hence this problem is difficult to solve.
A deep neural network (NN) has been used to find a practical solution to an inverse problem\cite{mccann_convolutional_2017} and can be used to restore the hidden structure in speckles~\cite{rahmani_multimode_2018} and infer scattering properties~\cite{seesan_deep_2022}.
We hypothesize that a deep NN technique can be used to extract features related to polarization randomization in speckles and synthesize the appearance of the RPE, as in DOPU imaging.

In this paper, we propose an NN-based approach to synthesize the DOPU from standard OCT signals.
A deep NN was trained using OCT intensity images and their corresponding DOPU images.
The trained network then synthesized DOPU images from unseen OCT intensity images.
The network's performance was evaluated by a grader who counted the clinical findings obtained from each DOPU image.

\section{Methods}\label{sec:method}

We collected data using a prototype OCT device (Section~\ref{sec:data-collection}).
Both the DOPU and OCT intensity were generated (Section~\ref{sec:signal-processing}) and used to train (Sections~\ref{sec:dataset-training} and \ref{sec:training_process}) and evaluate (Section~\ref{sec:evaluation-method}) a deep NN for synthesizing DOPU images (Section~\ref{sec:neural-network}).
The details of each step are as follows.

\subsection{Architecture of the syn-DOPU network}\label{sec:neural-network}

\begin{figure}
	\centering
	\includegraphics{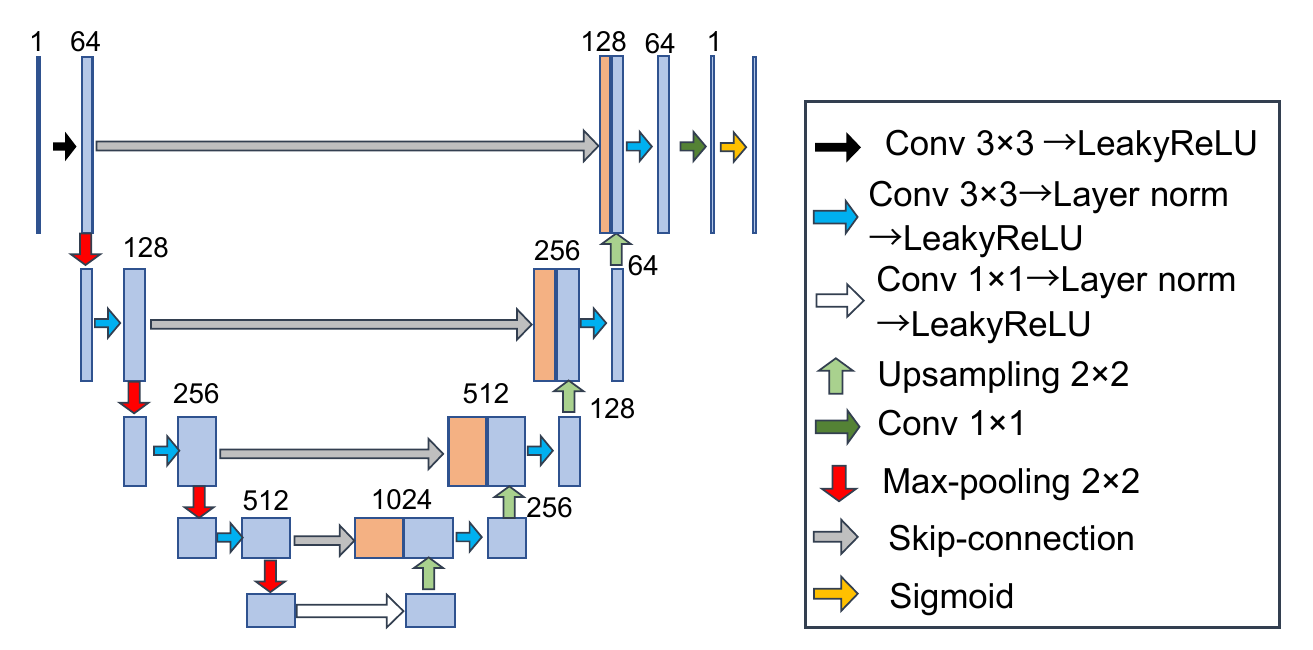}
	\caption{
		Neural network architecture for synthesizing the DOPU\@.
	}\label{fig:CNN_architecture}
\end{figure}

The network used in this study is based on the U-Net\cite{ronneberger_u-net_2015}, and its architecture is presented in Fig.~\ref{fig:CNN_architecture}.
The input OCT intensity first passes a 3\(\times\)3 convolution layer, and the channels are increased to 64.
In the encoder, the sequence of 3\(\times\)3 convolution, layer normalization\cite{ba_layer_2016}, and Leaky rectified linear unit (ReLU) activation function\cite{maas_rectifier_2013} (with a slope for negative values of 0.01) is applied followed by 2\(\times\)2 max-pooling.
The input size is halved by the operation while the number of channels is doubled.
This successive operation is applied four times in total.
At the bottleneck, the sequence of 1\(\times\)1 convolution, layer normalization, and Leaky ReLU activation is applied.
In the decoder, 2\(\times\)2 upsampling is applied to reconstruct the output from the extracted features.
Then, the output is concatenated with the corresponding output from the encoder part via a skip-connection path.
The sequence of 3\(\times\)3 convolution, layer normalization, and Leaky ReLU activation is applied.
The number of channels is reduced by a quarter except in the last stage.
A 1\(\times\)1 convolution is applied to reduce the number of channels from 64 to 1.
Finally, the sigmoid function is applied to limit the network's output within the range [0, 1], which is the range of realizable DOPU values.
The 1\(\times\)1 convolution is used at the bottleneck and the last stage of the decoder instead of a fully connected layer.
This is to fully connect all channels without spatial connections.
The network accepts variable input sizes larger than 16\(\times\)16.

\subsection{Data collection}\label{sec:data-collection}

A clinical-prototype polarization-diversity (PD)-OCT developed by us was used in this study\cite{makita_clinical_2018}.
This system does not have precise control of the polarization state of the probe beam, and only a compact polarization-diversity receiver (PDR) was installed.
In contrast to a standard PS-OCT system, this system cannot measure phase retardation; instead, it only measures DOPU with the minimum setup.

Briefly, the light source is a wavelength-swept laser (the central wavelength of 1 \um with a sweep rate of 100 kHz and the sweep range is approximately 100 nm).
The backscattered light is combined with the reference light, and finally, two orthogonal polarization components of the backscattered light are acquired by two detection channels of the PDR\@.
The sensitivity with respect to each channel is 89.5 dB.
The axial resolution is 6 \um in tissues defined by the full-width at half maximum, and the lateral resolution is expected to be 20 \um from a \(1/\rm{e}^2\) beam spot diameter.

Human eyes were scanned using the raster scan protocols (6 mm with 512 A-lines and 3 mm with 300 A-lines).
Four repeated B-scans were obtained at the same location.
Thirty-one volunteers without ophthalmic abnormalities (43 eyes) and 351 patients (478 eyes) were scanned.
The study was approved by the Institutional Review Boards of Tokyo Medical University and adhered to the tenets of the Declaration of Helsinki.
The nature of the present study and the implications of participating in this research project were explained to all study participants, and written informed consent was obtained from each participant before any study procedures or examinations were performed.

\subsection{Signal processing}\label{sec:signal-processing}

The DOPU and OCT intensity were obtained from the two OCT signals of the orthogonal polarization components of the backscattered probe beam.
DOPU images were used for the target of the NN training and the ground truth of the evaluation.
For input of the NN, OCT intensity corresponding to single-channel conventional OCT were obtained.

A modified DOPU reconstruction algorithm~\cite{makita_degree_2014} was used.
In this algorithm, the noise-bias-corrected Stokes vector is first calculated as follows:
\begin{equation}
	\label{eq:stokes_vector}
	\begin{bmatrix}
		s'_0(x,z,f) \\
		s'_1(x,z,f) \\
		s'_2(x,z,f) \\
		s'_3(x,z,f)		
	\end{bmatrix} =
	\begin{bmatrix}
		|g_\mathrm{H} (x,z,f)|^2 + |g_\mathrm{V}(x,z,f)|^2 -[n_\mathrm{H}(z)+n_\mathrm{V}(z)]\\
		|g_\mathrm{H} (x,z,f)|^2 - |g_\mathrm{V}(x,z,f)|^2 -[n_\mathrm{H}(z)-n_\mathrm{V}(z)]\\
		2\mathrm{Re}[g_\mathrm{H}(x,z,f) g_\mathrm{V}^{*}(x,z,f)] \\
		2\mathrm{Im}[g_\mathrm{H}(x,z,f) g_\mathrm{V}^{*}(x,z,f)]
	\end{bmatrix},
\end{equation}
where \(g_\mathrm{H}(x,z,f)\) and \(g_\mathrm{V}(x,z,f)\) are complex OCT signals obtained from the \(H\)- and \(V\)-polarization channels, \(x, z\) denotes the lateral and axial axes, \(f\) denotes the frame number taken at the same position, \(\mathrm{Re}\) and \(\mathrm{Im}\) are real and imaginary operators, respectively, and \(n_\mathrm{H}(z)\) and \(n_\mathrm{V}(z)\) denote the standard deviation of the background data in each channel.
Here, the \(H\)- and \(V\)-polarization channels detect the horizontally and vertically polarized components of light reaching the detector module, respectively.
From the Stokes vector [Eq.~(\ref{eq:stokes_vector})], the DOPU is calculated from \(N\) frames as
\begin{equation}
	\label{eq:def_DOPU}
	\mathrm{DOPU} (x,z) = \frac{\sum\limits^{N}_{f=1} \sqrt{\sum\limits^{3}_{m=1} \overline{s'_m}^2 (x,z,f)}}{\sum\limits^{N}_{f=1} \overline{s'_0}(x,z,f)},
\end{equation}
where \(\overline{s'_m}\) is the spatially averaged \(m\)-th Stokes parameter.
In this study, a moving average with 3\(\times\)3 pixels was used.

The corresponding OCT intensity is obtained from the coherent composition of each channel and complex-averaging of repeated scans\cite{ju_advanced_2013}, and then the absolute values are calculated as
\begin{equation}
	\label{eq:def_OCT_intensity}
	I(x,z) = \left| \frac{1}{N}\sum^{N}_{f=1} e^{-i \Delta\phi_f(x, z f)} \left[g_\mathrm{H}(x, z, f) + e^{-i \Delta\phi_\mathrm{ch}(x, f)} g_\mathrm{V}(x, z, f)\right] \right|^2,
\end{equation}
where \(\Delta\phi_f\) and \(\Delta\phi_\mathrm{ch}\) are the mean phase difference between the frames and channels, respectively.
This step can be thought of as mimicking the complex fields of the probe and reference beams before splitting them into two orthogonal polarization components by the PDR and taking the OCT signal composed of them.
This OCT signal corresponds to an elliptical polarization component of the backscattered probe beam.
This is equivalent to the typical commercial retinal OCT devices, which use a single-mode fiber-optic interferometer and a polarization controller to optimize OCT intensity.
Equation~(\ref{eq:def_OCT_intensity}) is corresponding to a single-channel OCT with a variable retarder whose axis is oriented horizontally or vertically in the reference arm.
Please see the details in \appendixname~\ref{sec:1ch-emulation}.
The complex averaging of repeated frames corresponds to a longer integration time.
Hence, this OCT intensity is essentially a conventional single-channel OCT intensity.

\subsubsection{Pre-processing for the NN}\label{sec:preprocess-network}

The DOPU is defined within the range [0, 1], and the network (Section~\ref{sec:neural-network}) output is within this range.
However, the DOPU [Eq.~(\ref{eq:def_DOPU})] with noise-bias-corrected Stokes vectors [Eq.~(\ref{eq:stokes_vector})] exceeds this range in no signal and low signal-to-noise ratio (SNR) regions because of the stochastic nature of noise.
This is because the ideal DOPU value without signal is indefinite [DOPU = 0/0].
The following process was used to handle invalid DOPU values in the DOPU images\cite{makita_degree_2014}:
\begin{equation}
	\label{eq:DOPU_prepro}
	\mathrm{DOPU}'(x,z)=
	\begin{cases}
		\mathrm{DOPU}(x,z) & 0<\mathrm{DOPU}(x,z)<1 \\
		1 & \text{otherwise}
	\end{cases}.
\end{equation}
Note that the pre-processing does not affect the DOPU values in regions where the OCT signal is larger than the background level and DOPU is not close to 0 or 1, i.e., at the scattering tissues exhibiting substantial polarization randomization within the penetration depth.
These are the areas of interest to be highlighted by DOPU\@.

As for NN input, OCT intensity is converted into SNR as
\begin{equation}
	\label{eq:OCT_prepro}
	I'(x,z) = (I(x,z) - \mu) / \sigma,
\end{equation}
where \(\mu{}\) is the background offset and \(\sigma{}\) is the standard deviation of the background.

\subsection{Dataset for training}\label{sec:dataset-training}

Twenty-six healthy volunteers (38 normal eyes) and 330 patients (457 pathological eyes) were included in the training and validation datasets.
The pathological cases included 145 patients of age-related macular degeneration (AMD), 42 patients of central serous chorioretinopathy (CSR), 22 patients of Vogt-Koyanagi-Harada disease (VKH), 17 patients of retinal vein occlusion, 13 patients of diabetic macular edema, 13 patients of myopia, and others.
Three pairs of OCT intensity and DOPU B-scans were extracted from each volume.
Then, multiple 64\(\times\)64-pixel patches in which less than 85\% of the patch had an SNR of less than 5 dB were extracted from each B-scan.

\subsection{Training process}\label{sec:training_process}

\begin{figure}
	\centering\includegraphics{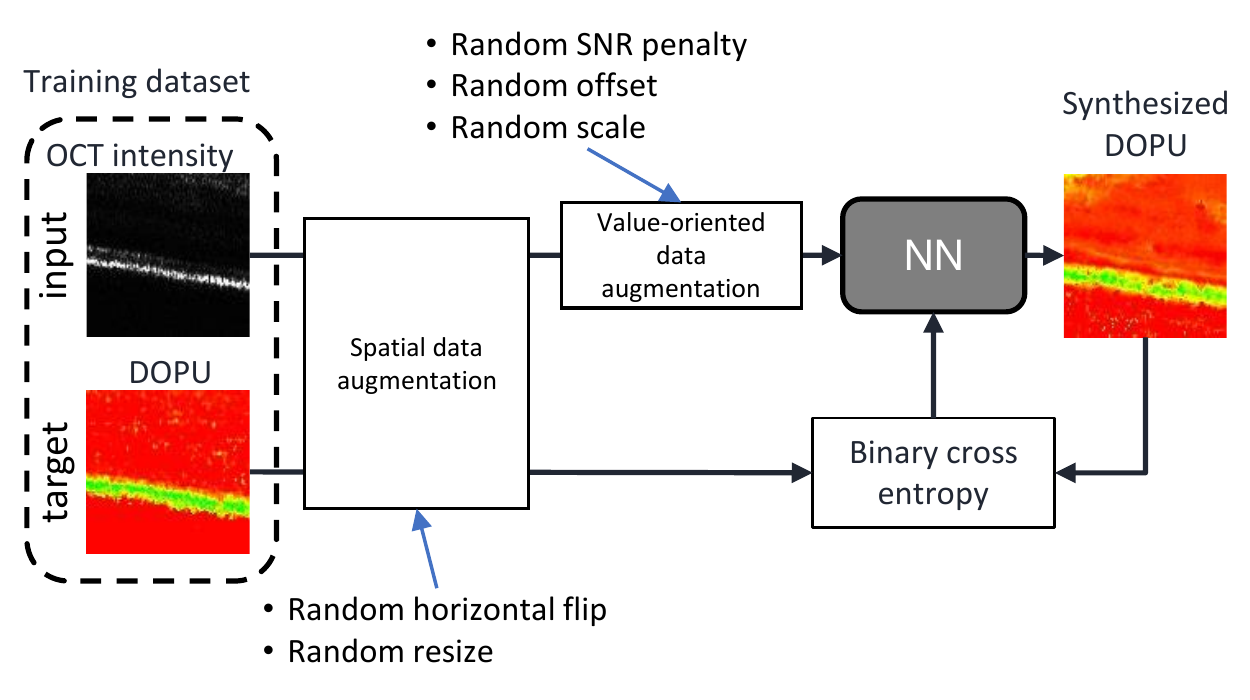}
	\caption{
		Training outline.
		A set of OCT intensity and DOPU in the dataset were randomly flipped in the horizontal direction and resized.
		For the OCT intensity values, random SNR penalizing, offset adding, and scaling are randomly applied, and then, the intensity was feed to the NN\@.
		The loss between the output of the NN and the target DOPU was obtained using binary cross entropy.
	} 
	\label{fig:outline_of_CNN_training}
\end{figure}

A diagram of the network training is shown in Fig.~\ref{fig:outline_of_CNN_training}.
The dataset was divided using a ratio of 8:2 by subject into training and validation datasets.
The total number of patches used for training was 69,478 (284 subjects), and 17,382 patches (72 subjects) were used for validation.

To increase the variety of the training data, the following data augmentations were applied during network training: horizontal flipping, resizing, SNR penalizing, offset adding, and scaling.
As shown in \figurename{}\ref{fig:outline_of_CNN_training}, spatial data augmentations were applied to both the input OCT intensity and target DOPU\@.
Both were randomly flipped in the horizontal direction and resized with magnifications of [1, 2] in the axial and [1, 4] in the horizontal directions.
Data augmentations for the OCT intensity values were as follows.
Random SNR penalizing from 0 to -15 dB, scaling from 0.01 to 100 times, and offset adding from 0 to 1,000 were applied.
Details of the SNR penalizing are described in \appendixname~\ref{sec:snr-penalty}.

Because the sigmoid function is used to bound the output range (Section~\ref{sec:neural-network}) and the target DOPU values are concentrated at the edge of the range, the gradient of the sigmoid output could be frequently saturated\cite{reed_neural_1999}.
To avoid a small gradient of error, the binary cross entropy\cite{bishop_pattern_2006} between the output of the network and the corresponding target DOPU was used as the loss function.
\begin{equation}\label{eq:loss-function}
	L_{\mathrm{BCE}} = \sum_{n, w, h} -\left[y_{n,w,h}\log(x_{n,w,h}) + (1 - y_{n,w,h}) \log(1 - x_{n,w,h})\right],
\end{equation}
where \(x\) is the output of the NN; \(y\) is the target DOPU; \(n, w, h\) are the indices of a pixel in a minibatch with a batch size of \(N\), width of \(W\), and height of \(H\).
The NN was trained using minibatch learning with a batch size of 64.
The Adam optimizer\cite{kingma_adam_2017} with a learning rate of 0.001 and momentum parameters \(\beta_1\) = 0.9 and \(\beta_2\) = 0.999 was used.
The training was stopped when the validation loss did not improve for 20 epochs.
The epoch that exhibited the minimum validation loss was used as the trained network.

\subsection{Evaluation of DOPU synthesis}\label{sec:evaluation-method}

We evaluated the trained DOPU-synthesizing network from the perspective of clinical utility.
The potential of DOPU would be to provide additional contrasts of RPE abnormalities.
To analyze the ability of syn-DOPU to reproduce abnormal appearances of the RPE, an abnormality-level detection was performed.
Agreements of the abnormality detection between the DOPU and syn-DOPU images were then calculated for normal and pathological eyes.
This abnormal appearance detection is analogous to object detection.
Hence, we apply an analysis similar to ``object count''\cite{wolf_object_2006}.

There are three types of abnormality findings:
the DOPU and syn-DOPU images both show the same abnormality at the same location [DOPU(+) $\land$ syn-DOPU(+)],
the DOPU image shows an abnormality, but the same location in the syn-DOPU image does not [DOPU(+) $\land$ syn-DOPU(-)],
and the syn-DOPU image shows an abnormality, but the same location in the DOPU image does not [DOPU(-) $\land$ syn-DOPU(+)].
Note that the case in which both the DOPU and syn-DOPU images do not show abnormalities is not available in this protocol because counts of normal regions cannot be defined similarly to object detection\cite{padilla_survey_2020}.
.
The numbers of these findings in each cross-sectional image were counted for each abnormality.
As same as object count\cite{wolf_object_2006}, recall and precision as measures of the abnormality reproduction performance were calculated as follows
\begin{equation}
	\label{eq:def_recall}
	\rm{Recall} = \frac{\#[\textrm{DOPU(+)} \land \textrm{syn-DOPU(+)}]}{\#[\textrm{DOPU(+)} \land \textrm{syn-DOPU(+)}] + \#[\textrm{DOPU(+)} \land \textrm{syn-DOPU(-)}]}
\end{equation}
\begin{equation}
	\label{eq:def_precision}
	\rm{Precision} = \frac{\#[\textrm{DOPU(+)} \land \textrm{syn-DOPU(+)}]}{\#[\textrm{DOPU(+)} \land \textrm{syn-DOPU(+)}] + \#[\textrm{DOPU(-)} \land \textrm{syn-DOPU(+)}]},
\end{equation}
where \#[] is the count of each appearance.
From the Recall and Precision, F\(_1\) score can be calculated as:
\begin{equation}
	\label{eq:def_f1}
	\rm{F_1 score} = 2 \frac{\rm{Recall} \cdot \rm{Precision}}{\rm{Recall} + \rm{Precision}}.
\end{equation}

In the evaluation, we used the data of five volunteers (five normal eyes) and 20 patients (20 pathological eyes), which were not included in the training or validation datasets.
The pathological eyes included 12 AMDs, 5 CSRs, 2 pachychoroids, and 1 case of myopia.
The DOPU and synthesized DOPU (syn-DOPU) images of the evaluation dataset were evaluated by an ophthalmologist.
The protocols of the RPE-abnormality detection for normal and pathologic cases were as follows.

\subsubsection{RPE-abnormality detection in normal eyes}

In each eye, five B-scans equally spaced along the slow scan direction were extracted.
Then, the ophthalmologist evaluated the DOPU and syn-DOPU images to find abnormal appearances in the RPE\@.
Cross-sectional images of OCT intensity and two DOPU images were presented simultaneously to evaluate the RPE abnormalities.

\subsubsection{RPE-abnormality detection in pathological eyes}\label{sec:eval-method_pathologic}

For pathologic eye cases, abnormal appearances were detected by the ophthalmologist in DOPU and syn-DOPU images these were selected to include abnormal regions.
First, another ophthalmologist viewed the OCT intensity volume data and selected the pathological region of abnormality along the slow scanning direction for each volume.
Then, five B-scans were randomly selected from each region.
The ophthalmologist then surveyed the DOPU and syn-DOPU images to demarcate four types of abnormalities: RPE defects, RPE thickening, RPE elevation, and intraretinal hyperreflective foci (HRF) considered as intraretinal RPE migration\cite{miura_evaluation_2017}.
The information of subjects was blinded.
Cross-sectional images of OCT intensity and two DOPU images were presented simultaneously to the grader, and the syn-DOPU image of the same locations where abnormalities were observed on the DOPU image, and vice versa, were carefully evaluated.

\section{Results}

\subsection{Normal eye}
\begin{figure}
	\centering\includegraphics{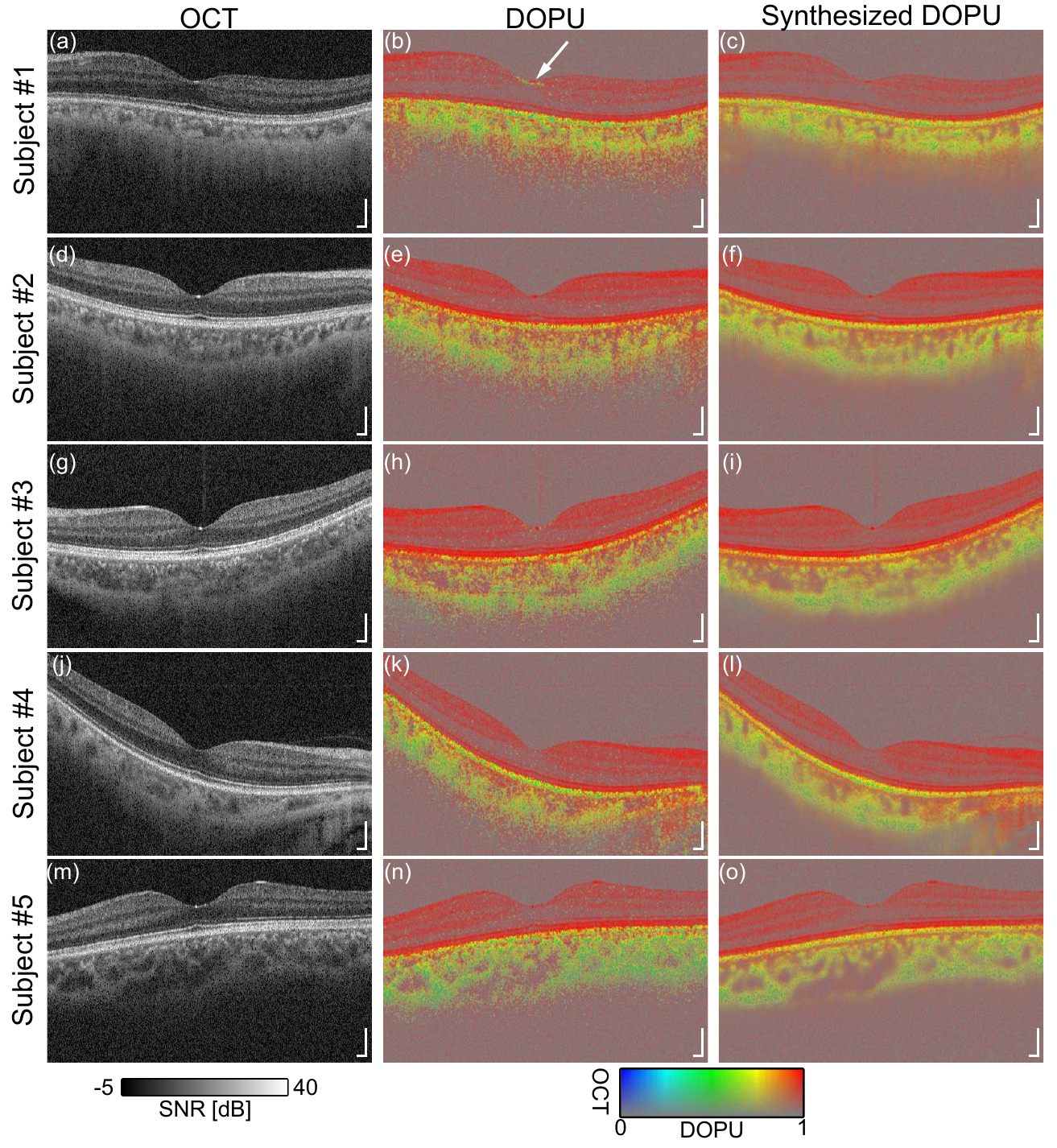}
	\caption{
Inference of syn-DOPU images for the five eyes of five normal subjects.
		(a, d, g, j, m) OCT, (b, e, h, k, n) DOPU, and (c, f, i, l, o) syn-DOPU cross-sectional images.
		Scale bars indicate 200 \um.
	}\label{fig:normal_samples}
\end{figure}

Figure~\ref{fig:normal_samples} shows the OCT intensity, DOPU, and syn-DOPU images of all 5 normal cases in the evaluation dataset.
For all eyes, the whole appearance of the DOPU is well reproduced in the syn-DOPU image.
No abnormal appearances were found in any of the 25 syn-DOPU or DOPU images.
Hence, we could not evaluate recall, precision, and F\(_1\) score.
Sometimes, the DOPU images exhibited a low DOPU value at the inner limiting membrane of the fovea (white arrow in \figurename{}~\ref{fig:normal_samples}).
This appearance is not reproduced in the corresponding syn-DOPU image.
The NN generated a normal distribution of DOPU in the syn-DOPU images.

\subsection{Pathologic cases}\label{sec:evaluate-result-pathology}
\begin{figure}
	\centering\includegraphics{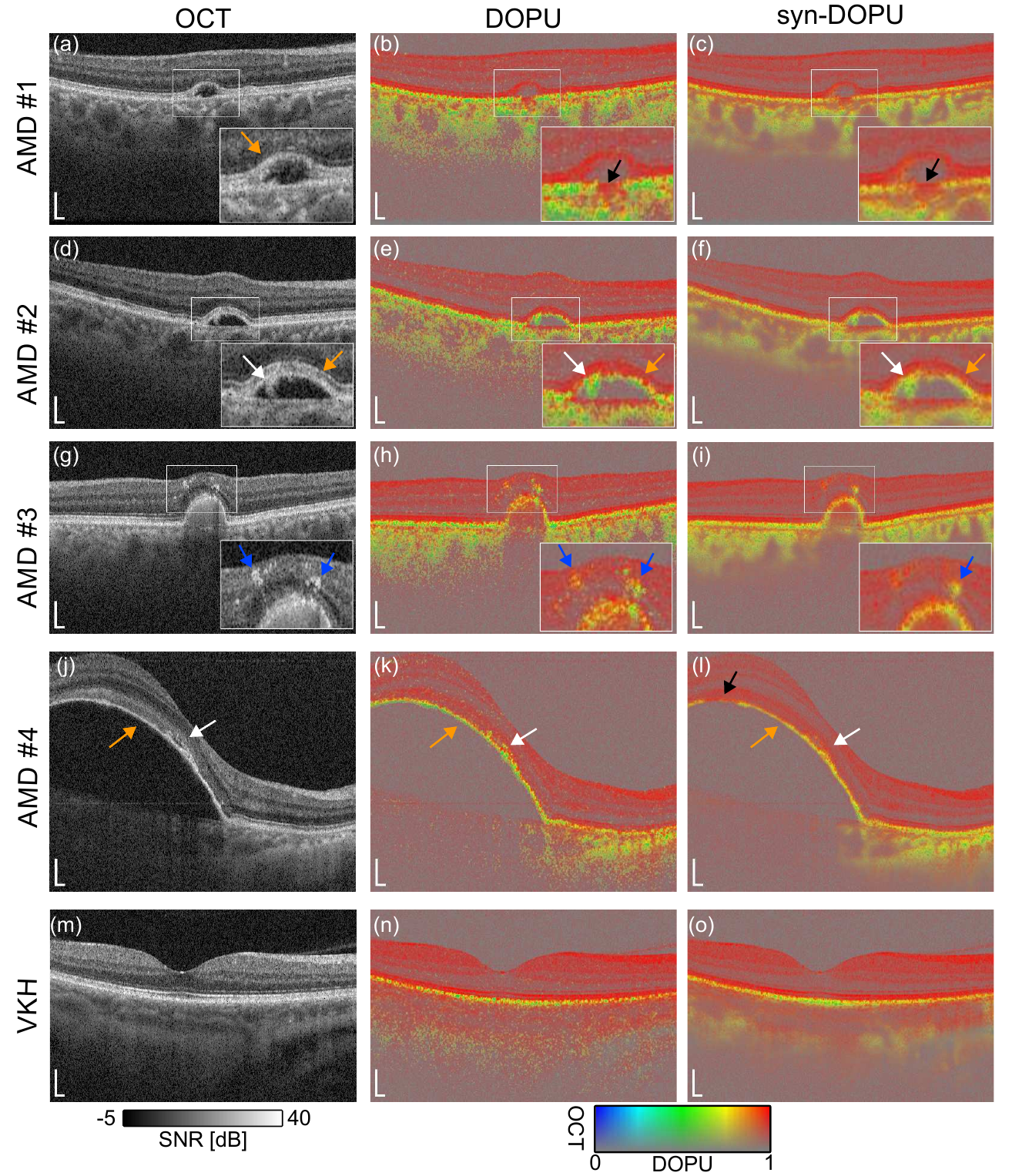}
	\caption{
Inference of the syn-DOPU for 5 eyes of 5 pathological cases.
		(a, d, g, j, m) OCT intensity, (b, e, h, k, n) DOPU, and (c, f, i, l, o) syn-DOPU cross-sectional images.
		Colored arrows indicate the four types of RPE abnormalities: RPE defects (black), RPE elevation (orange), RPE thickening (white), and HRF (blue).
		For the VKH case (m-o), both the DOPU and syn-DOPU images exhibit high DOPU values in the choroid.
		Scale bars indicate 200 \um.
	}\label{fig:pathological_samples}
\end{figure}

Figure~\ref{fig:pathological_samples} shows the OCT intensity, DOPU, and syn-DOPU images of 5 pathological cases.
AMD \#1 exhibits detachment of the hyper-scattering layer in the OCT intensity image [orange arrow, \figurename{}~\ref{fig:pathological_samples}(a)] where the DOPU values are high in both the DOPU and syn-DOPU images [\figurenames{}~\ref{fig:pathological_samples}(b) and \ref{fig:pathological_samples}(c)].
Hence, the hyper-scattering layer might be photoreceptors, and this will be serous retinal detachment.
RPE defects were identified in both the DOPU and syn-DOPU images (black arrows), where apparent abnormalities at the RPE are not observed in the OCT intensity image.
In AMD \#2, detachment of the hyper-scattering layer in OCT intensity image is visible (orange arrow), which was identified as RPE elevation in both DOPU and syn-DOPU images.
In addition, the thickening of the hyper-scattering layer appeared in the OCT intensity image, and it was graded as RPE thickening in both the DOPU and syn-DOPU images (white arrow).
The RPE seems elevated in both the AMD \#1 and \#2 OCT intensity images (orange arrows).
However, the DOPU image of AMD \#1 [\figurename{}~\ref{fig:pathological_samples}(b)] does not show low DOPU signals, whereas that of AMD \#2 [\figurename{}~\ref{fig:pathological_samples}(e)] does (graded as an RPE elevation, orange arrows).
This tendency is correctly reproduced in the syn-DOPU images [\figurenames{}~\ref{fig:pathological_samples}(c) and \ref{fig:pathological_samples}(f)].
This indicates the potential ability of the proposed method to generate biomarkers for RPE abnormalities obtained by the DOPU without a PS-OCT setup.

In AMD \#3, HRFs are shown in both the DOPU and syn-DOPU images [blue arrows, \figurenames{}~\ref{fig:pathological_samples}(h) and \ref{fig:pathological_samples}(i)].
However, some HRFs do not appear in the syn-DOPU image.
In AMD \#4, RPE elevation was observed by the grader in the DOPU image [white arrow, \figurename{}~\ref{fig:pathological_samples}(k)].
An RPE defect was observed in the syn-DOPU image [black arrow, \figurename{}~\ref{fig:pathological_samples}(l)] that was not observed in the DOPU image.

\begin{table}[]
	\caption{Summary of the counts of clinical findings observed from the DOPU and syn-DOPU images.
	D(\(\pm\)): DOPU positive or negative, SD(\(\pm\)): Syn-DOPU positive or negative.}\label{table:Summary-of-abnormalities-findings}
	\centering
	\begin{tabular}{l|cccccc}
		& \multicolumn{3}{c}{\# of findings} & \multirow{2}{*}{Recall} & \multirow{2}{*}{Precision} & \multirow{2}{*}{F\(_1\) score} \\
		& \makecell{D (+)\\ \(\land\) \\ SD (+)} & \makecell{D (+)\\ \(\land\) \\ SD (-)} & \makecell{D (-)\\ \(\land\) \\ SD (+)} & & & \\
		\midrule
		RPE defect     & 32 &  4 & 11 & 0.889 & 0.744 & 0.810 \\
		RPE thickening & 22 &  3 &  0 & 0.88 & 1 & 0.936 \\
		RPE elevation  & 63 &  0 &  0 & 1 & 1 & 1 \\
		HRF            &  9 & 12 &  0 & 0.429 & 1 & 0.600 \\
		Total		   &126 & 19 & 11 & 0.869 & 0.920 & 0.894
	\end{tabular}
\end{table}

The findings of the RPE abnormalities in the DOPU and syn-DOPU images are summarized in Table~\ref{table:Summary-of-abnormalities-findings}.
The recalls of each abnormality were 0.889 (RPE defects), 0.88 (RPE thickening), 1 (RPE elevation), and 0.429 (HRF).
The precisions of each abnormality were 0.744 (RPE defects), 1 (RPE thickening), 1 (RPE elevation), and 1 (HRF).
The F\(_1\) score of each abnormality were 0.810 (RPE defects), 0.936 (RPE thickening), 1 (RPE elevation), and 0.600 (HRF).
The recall, precision, and F\(_1\) score of all four abnormalities are 0.869, 0.920, and 0.894, respectively.
Overall, the findings of the RPE abnormalities are well reproduced by syn-DOPU\@.

In addition, a VKH case was evaluated.
As VKH progresses, choroidal melanin is reduced\cite{okeefe_vogt-koyanagi-harada_2017} and choroidal DOPU values are increased~\cite{miura_polarization-sensitive_2017,miura_objective_2022}
The syn-DOPU image of the VKH case [\figurename{}~\ref{fig:pathological_samples}(o)] shows high DOPU values at the choroid, as in the DOPU image [\figurename{}~\ref{fig:pathological_samples}(n)].
The NN hence could be able to also generate biomarkers for choroidal pigmentation in syn-DOPU images as well.

\section{Discussion}

\subsection{Signal source of the syn-DOPU}

The proposed network was trained to synthesize the DOPU from OCT intensity images corresponding to one polarization component, even though the real DOPU is calculated from two polarization components obtained by PS-OCT\@.
The input to the network lacks sufficient degrees of freedom to determine the polarization state.
Hence, the factors that generate the features in DOPU images probably also influence OCT intensity.

One factor is scattering properties\cite{bicout_depolarization_1994,morgan_effects_1997,mishchenko_depolarization_1995}.
One main source of DOPU contrast in retinal imaging is considered to be melanin\cite{gotzinger_retinal_2008-1,baumann_polarization_2012, baumann_melanin_2015}.
Although the effect of polarization scrambling is only measured by PS-OCT, different scattering properties affect the speckle formation in OCT intensity images\cite{karamata_multiple_2005-1,hillman_detection_2010}.
The melanin will also cause wavelength-dependent attenuation, and it is found to be correlated with DOPU\cite{merkle_degeneration_2022}.
Although it is unclear whether it is a source of the DOPU or not, it might also affect the speckle formation.
The NN could be synthesizing the DOPU images from these features in the OCT intensity images.

Another factor might be the morphological patterns that appear in the OCT-intensity images.
Some parts of the spatial pattern, such as the RPE and choroid, are quite similar in both the OCT intensity and DOPU images.
The U-Net might utilize the similarity of these spatial structures when synthesizing DOPU images.

\subsection{Performance and challenges of the proposed method}

Normal and pathological eyes were used to evaluate the syn-DOPU network.
The evaluation was conducted to detect abnormal appearances at the RPE\@.
In the case of pathological eyes, the evaluation was conducted to check whether the NN can correctly reproduce 4-type RPE abnormalities without missing abnormalities or synthesizing false ones.

In the normal eye cases, there were no abnormalities in the 25 syn-DOPU (5 eyes) evaluation images.
The proposed method faithfully synthesizes normal appearances.
For the pathological eye cases, recall and precision were computed for each abnormality.
The recall reveals how often this method reproduces abnormalities that appear in the DOPU images, whereas the precision reveals how many synthesized abnormalities agree with the appearances in DOPU images.
The overall recall and precision of the RPE abnormalities are high (>0.85).

The precision for RPE defects (0.744) is lower than that of other abnormalities (1.0).
When some hyper-scattering materials surround the RPE, as shown in AMD \#4 in \figurename{}~\ref{fig:pathological_samples}, false RPE defects appear.
Because the U-Net uses the spatial structure, the decreased contrast between the RPE and surrounding tissue might be a source of false defects.

The recall was especially low for HRF (0.429).
Some HRFs are not well generated by syn-DOPU\@.
This is perhaps due to the small size of the HRFs.
In both the training and validation datasets, the HRF region can be very small.
Increasing the HRF cases in the training dataset and weighting the abnormal regions during training will improve the synthesis of HRF\@.

\subsection{Limitation in the evaluation of abnormal appearances}\label{sec:discuss-limit-eval}

\begin{figure}
	\centering
	\includegraphics{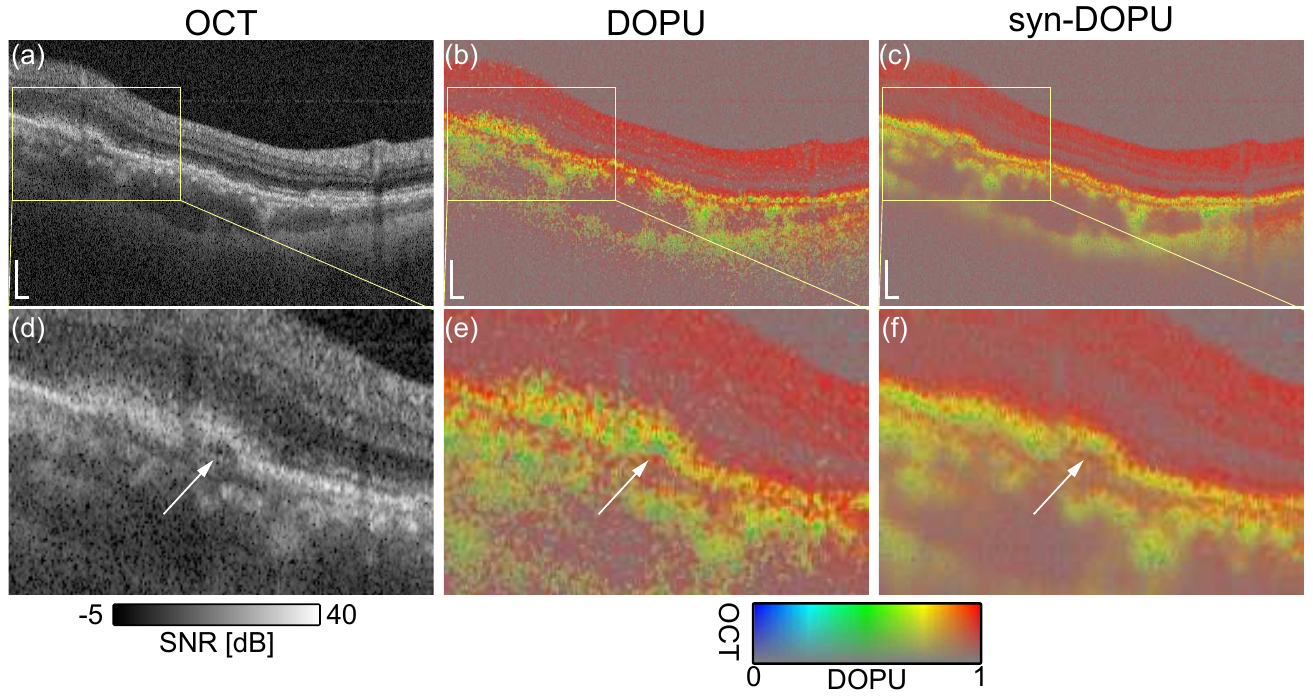}
	\caption{A case of Syn-DOPU shows abnormality better than DOPU\@.
	(a) OCT, (b) DOPU, and (c) syn-DOPU cross-sectional images of AMD \#4 at a different scanning location to \figurename{}~\ref{fig:pathological_samples}(j-l).
	A sub-RPE void appeared in the OCT cross-sectional image (d) and can be confirmed in the DOPU image (f), however, is not clear in the DOPU image (e).
	Scale bars indicate 200 \um.}\label{fig:PED-betterSynDOPU}
\end{figure}

We analyzed the detection of abnormal appearances in DOPU and syn-DOPU images.
An issue in the evaluation is that the false positive [DOPU(-) $\land$ syn-DOPU(+)] in this analysis does not necessarily mean less utility of syn-DOPU\@.
The evaluation is the comparison between image appearances.
Thus, normal appearances in DOPU do not always mean not abnormal and vice versa.
Some false positives may indicate higher sensitivity of syn-DOPU more than DOPU\@.
As shown in \figurename{}~\ref{fig:PED-betterSynDOPU}, the ophthalmologist found some DOPU images do not show abnormalities well where syn-DOPU does.
There is a void beneath the RPE in the OCT intensity image [white arrow in \figurename{}~\ref{fig:PED-betterSynDOPU}(d)], and it can be recognized in the syn-DOPU image [\figurename{}~\ref{fig:PED-betterSynDOPU}(f)].
However, the DOPU image shows that most areas of the void have low DOPU values [\figurename{}~\ref{fig:PED-betterSynDOPU}(e)].
The DOPU imaging should depend on several factors such as the incident polarization state\cite{lippok_quantitative_2019} and the size\cite{sugita_analysis_2015} and shape\cite{hsu_quantitative_2020} of the averaging kernel of Stokes vector.
Hence, it is not guaranteed that DOPU images can always show clear RPE abnormalities.
The neural network has been trained on DOPU images in a variety of conditions, so it may already be generalized to some of these factors.
This might be the reason why some syn-DOPU images visualize abnormalities better than DOPU\@.
We cannot evaluate which is better in the current study.
To assess this, it would require a longitudinal study to follow several patients and compare DOPU and syn-DOPU images with clinical outcomes, such as visual acuity, size of abnormalities, etc.

\subsection{Treatment of noise values in the DOPU}\label{sec:DOPU_prepro}

If there is no reflection from the measurement sample, the DOPU is not correctly reconstructed.
In the case of the human eye, signals deeper than the choroid cannot be well measured.
After applying depth-dependent noise correction\cite{makita_degree_2014}, the signal will appear as noise that takes a value greater than 1 or less than 0.
To ignore this noise during training, pre-processing [Eq.~(\ref{eq:DOPU_prepro})] was applied.

There could be other pre-processing methods to handle this problem.
The utility of the used pre-processing method [Eq.~(\ref{eq:DOPU_prepro})] has been evaluated by comparing it with others.
Two other pre-processing methods were defined:
\begin{equation}
	\label{eq:DOPU_prepro_0_0}
	M_{0, 0}[x]=
	\begin{cases}
		x & 0<x<1 \\
		0 & \text{otherwise}
	\end{cases}
\end{equation}
and
\begin{equation}
	\label{eq:DOPU_prepro_0_1}
	M_{0, 1}[x]=
	\begin{cases}
		x & 0<x<1 \\
		1 & x \geq 1 \\
		0 & x \leq 0
	\end{cases}
	.
\end{equation}

\begin{figure}
	\centering
	\includegraphics{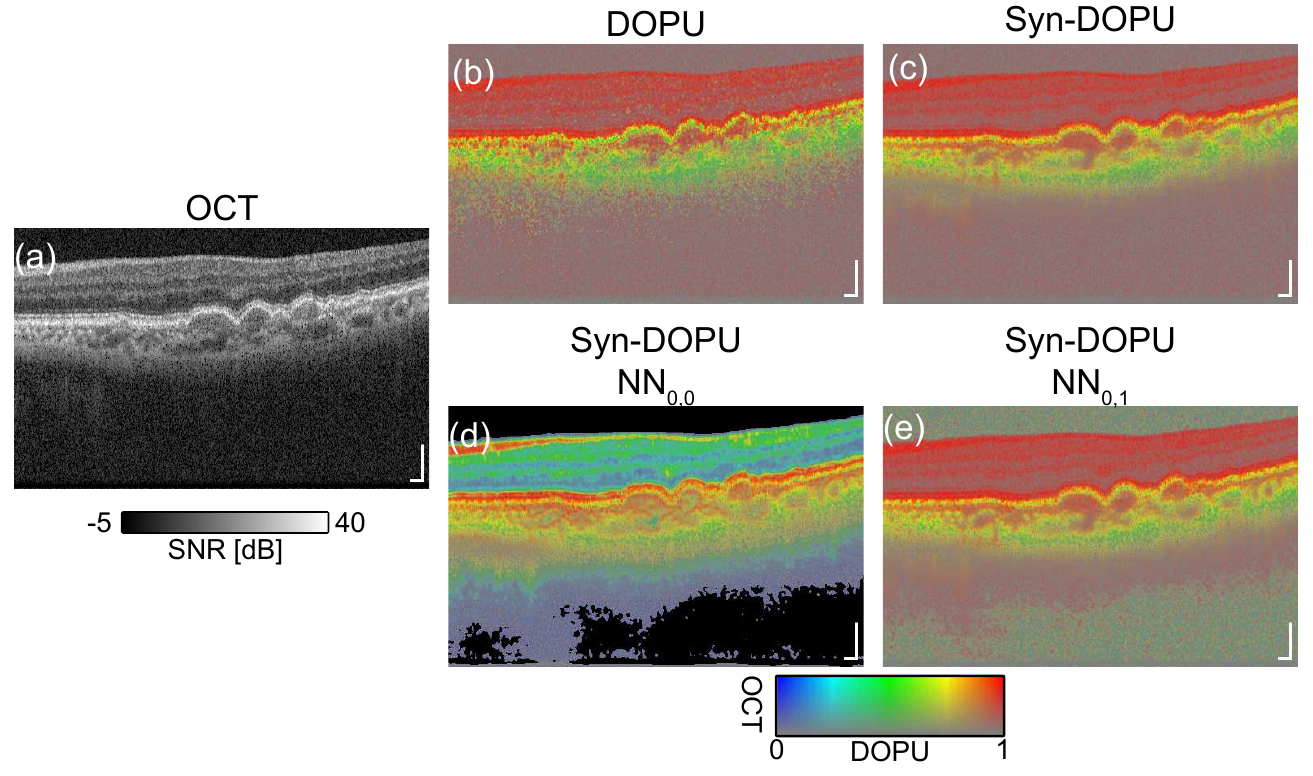}
	\caption{
Syn-DOPU images generated by networks trained using different pre-processing methods for the target DOPU\@.
		(a) OCT, (b) DOPU, and (c--e) syn-DOPU images of a subject in the evaluation dataset.
		Three syn-DOPU images, (c--e), were obtained from three networks trained using Eqs.~(\ref{eq:DOPU_prepro}), (\ref{eq:DOPU_prepro_0_0}), and (\ref{eq:DOPU_prepro_0_1}), respectively, as pre-processing for the target DOPU\@.
		Scale bars indicate 200 \um.
	}
	\label{fig:DOPU-prepro}
\end{figure}

Next, the performance results of pre-processing using Eqs.~(\ref{eq:DOPU_prepro}), (\ref{eq:DOPU_prepro_0_0}), and (\ref{eq:DOPU_prepro_0_1}) were compared.
The NNs trained using Eqs.~(\ref{eq:DOPU_prepro_0_0}) and (\ref{eq:DOPU_prepro_0_1}) as pre-processing are referred to as NN\(_{0,0}\) and NN\(_{0,1}\), respectively.

The resulting syn-DOPU images are shown in Fig.~\ref{fig:DOPU-prepro}.
A subject from the evaluation dataset was used.
For NN\(_{0,0}\) [\figurename{}~\ref{fig:DOPU-prepro}(d)], the DOPU values in the retina are too low.
This may be because the network was trained to generate low DOPU values in the lower SNR regions.
For NN\(_{0,1}\) [\figurename{}~\ref{fig:DOPU-prepro}(e)], the DOPU values in the tissue regions are reasonable.
In the noise regions, the DOPU values are around 0.5; this is the expected DOPU value (the middle of [0, 1]).

The root-mean-squared error (RMSE) between the DOPU and syn-DOPU in low SNR regions (between 1 and 5 dB) was calculated using the evaluation dataset (25 volumes of 20 pathological and 5 healthy eyes).
If Eq.~(\ref{eq:DOPU_prepro}) is used, the RMSE is 0.044\(\pm\)0.004 (mean\(\pm\)STD over 25 volumes), whereas NN\(_{0,0}\) and NN\(_{0,1}\) yield RMSEs of 0.926\(\pm\)0.008 and 0.269\(\pm\)0.014, respectively.
Although the syn-DOPU image obtained from NN\(_{0,1}\) [\figurename{}~\ref{fig:DOPU-prepro}(e)] seems good, the syn-DOPU values at the low SNR pixels depart slightly from the DOPU values.
Given these results, we decided to use Eq.~(\ref{eq:DOPU_prepro}) for pre-processing.

\subsection{Variable input size}

\begin{figure}
	\centering
	\includegraphics{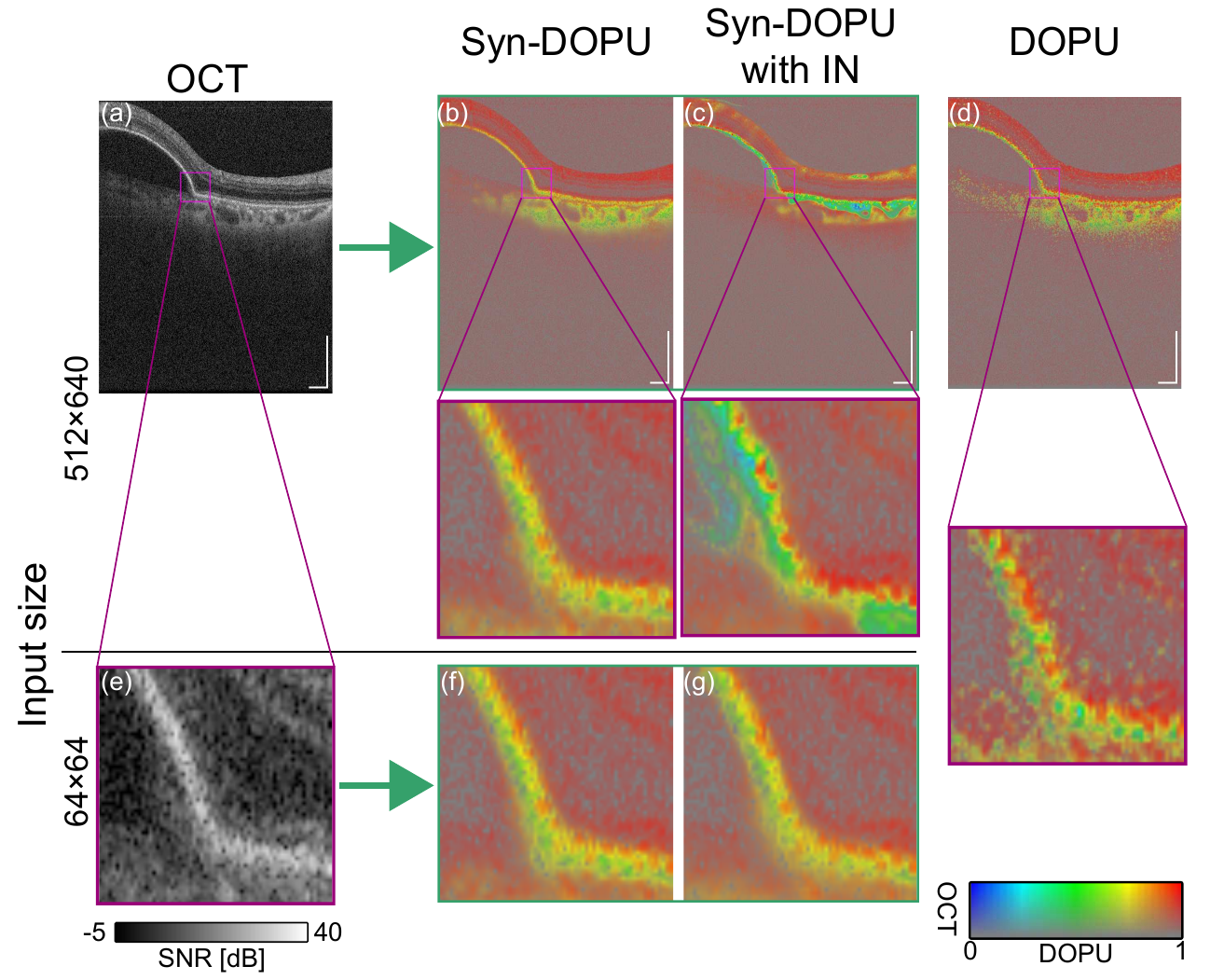}
	\caption{
		Comparison of the DOPU synthesized by the NN using layer normalization and instance normalization.
		When the input OCT intensity image is 512\(\times\)640 pixels (a), the DOPU synthesized by the proposed network (b) agrees well with the ground truth DOPU (d).
		However, the DOPU synthesized by a network with instance normalization (c) does not agree well with (d).
		If the input is cropped (e) to the size of training dataset images (64\(\times\)64 pixels), the syn-DOPU images obtained by both networks (f, g) agree well with the ground truth.
		Scale bars indicate 500 \um.
	}\label{fig:variable_input_size}
\end{figure}

To generate images, NNs frequently use instance normalization\cite{ulyanov_instance_2017}.
However, instance normalization is not suitable for the proposed method because the input size is variable.
We used layer normalization instead because it does not depend on data size\cite{ba_layer_2016}.

Figure~\ref{fig:variable_input_size} shows DOPU images synthesized from inputs of different sizes.
For comparison, a network with the architecture in \figurename{}~\ref{fig:CNN_architecture} except that the layer normalization layers were replaced by instance normalization layers was trained on the same training dataset (Section~\ref{sec:dataset-training}).
The network with instance normalization synthesized the DOPU well when the input size [64\(\times\)64 pixels, \figurename{}~\ref{fig:variable_input_size}(e)] is the same as that of the training dataset, as shown in \figurename{}~\ref{fig:variable_input_size}(g).
However, when the input size [512\(\times\)640 pixels, \figurename{}~\ref{fig:variable_input_size}(a)] differs from that of the training dataset, the syn-DOPU image [\figurename{}~\ref{fig:variable_input_size}(c)] exhibits discrepancies in the values and structure of the low-DOPU regions with respect to the ground truth [\figurename{}~\ref{fig:variable_input_size}(d)].
In contrast, the syn-DOPU for both input sizes are close to the ground truth when the proposed network is used [\figurenames{}~\ref{fig:variable_input_size}(b) and \ref{fig:variable_input_size}(f)].
The RMSEs of the DOPU and syn-DOPU for 512\(\times\)640 inputs were calculated with the evaluation dataset (25 volumes of 20 pathological and 5 healthy eyes).
The RMSEs in the high SNR regions (more than 1 dB) are 0.063\(\pm\)0.005 (mean\(\pm\)STD over 25 volumes) and 0.137\(\pm\)0.012 when NNs with layer normalization and with instance normalization are used, respectively.
The NN with layer normalization yields better inference accuracy.

\subsection{Transferring to non-PS-OCT}

\begin{figure}
	\centering
	\includegraphics{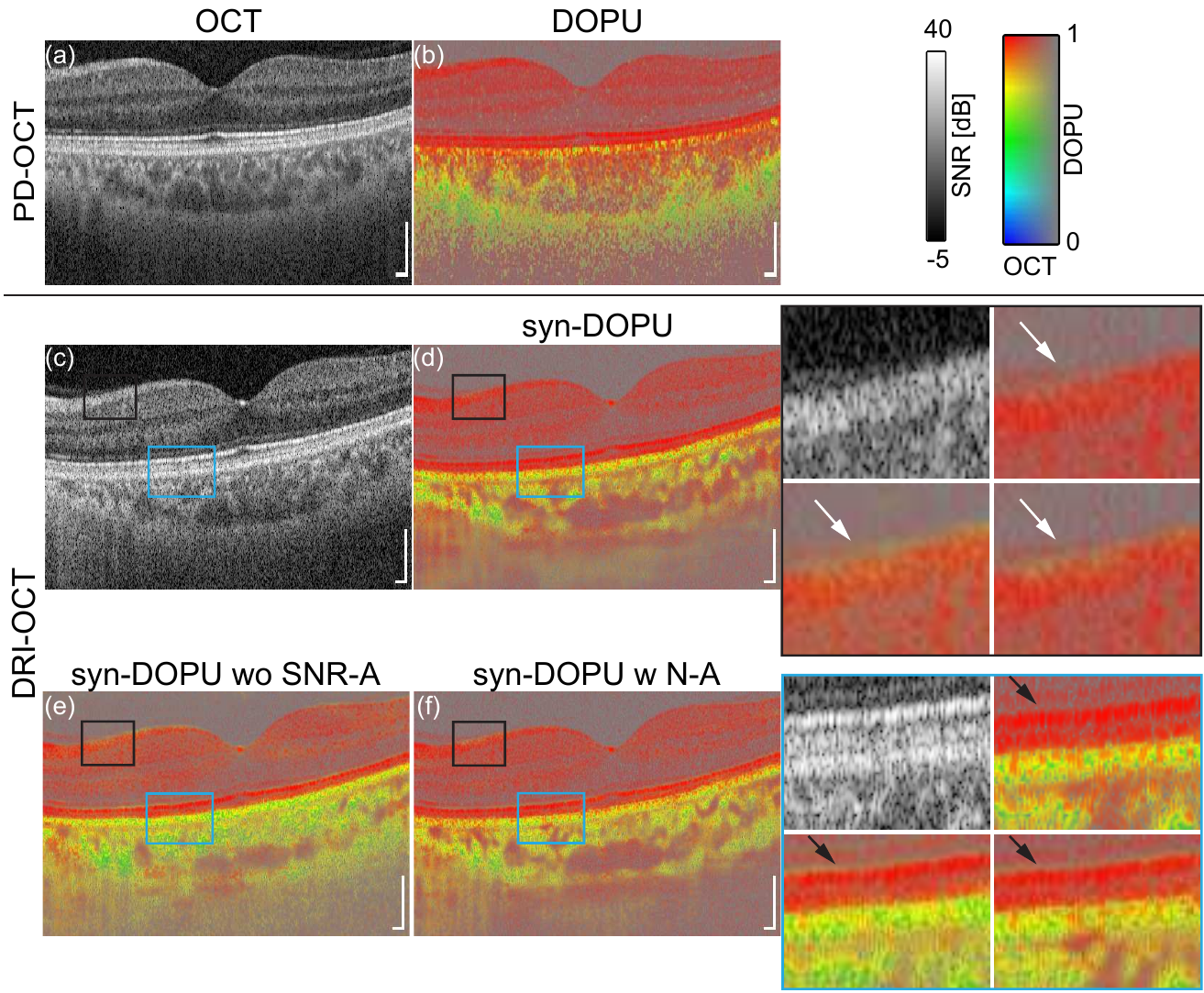}
	\caption{
		Syn-DOPU image obtained with a non-PS-OCT setup.
		The same healthy eye has been scanned with PD-OCT (a) and a commercial OCT device (c).
		The syn-DOPU obtained from the commercial OCT (d) shows no abnormalities, as does the ground truth DOPU (b) obtained by PD-OCT\@.
		The syn-DOPU results obtained by NNs trained (e) without random SNR penalizing and (f) with random exponential noise exhibit incorrectly low DOPU values at the retinal nerve fiber layer (white arrows) and photoreceptor layer (black arrows).
		Scale bars indicate 200 \um.
	}\label{fig:syndopu-drioct}
\end{figure}

In practice, the proposed method will be transferred to a non-PS OCT system.
The different imaging properties of the two systems will be challenges to overcome during implementation.
Figure~\ref{fig:syndopu-drioct} shows a typical normal case of the syn-DOPU obtained from a non-PS commercial OCT device (DRI-OCT, Topcon).
The spatial sampling properties of the image with the commercial OCT device are 11.7 \um/A-line and 2.6 \um/pixel.
The B-scan OCT intensity image is not averaged; hence, the SNR would be lower than that used in training (the complex values of the 4 B-scans are averaged as described in Section~\ref{sec:signal-processing}).
The syn-DOPU [\figurename{}~~\ref{fig:syndopu-drioct}(d)] exhibits a good appearance, similar to the DOPU obtained by PD-OCT [\figurename{}~~\ref{fig:syndopu-drioct}(b)], with high DOPU values in the retina and low DOPU values at the RPE and choroid.
Network training with value-oriented data augmentation (Section~\ref{sec:training_process}) may be necessary to handle OCT intensity images from different setups.
In this study, we introduced the SNR penalizing, which emulates model-based additional additive random noise in OCT intensity images (\appendixname~\ref{sec:snr-penalty}), as one data augmentation.
The effect of the SNR augmentation is shown by comparing syn-DOPU results obtained without SNR augmentation [\figurename{}~\ref{fig:syndopu-drioct}(e)] and with noise augmentation [\figurename{}~\ref{fig:syndopu-drioct}(f)].
The noise augmentation adds a random variable that obeys an exponential distribution to the OCT intensity, i.e., Eq.~(\ref{eq:snr-penalty-intensity}) without the cross term between the signal and noise.
When SNR augmentation is not used, the DOPU values in the syn-DOPU images are too low at the boundary between the vitreous and retina (white arrows) and at the inner/outer segment junction (black arrows).
Hence, proper data augmentations are required to generalize the method for other different devices.

In the choroid, it seems the syn-DOPU values [\figurename~\ref{fig:syndopu-drioct}(d)] are slightly lower than those of DOPU obtained by PD-OCT [\figurename~\ref{fig:syndopu-drioct}(b)].
This is maybe because the different numbers of B-scans are used.
For the training, we used a multi-frame complex averaged OCT signal [Eq.~(\ref{eq:def_OCT_intensity})] as the input to the network.
In the case of the commercial OCT device [\figurenames~\ref{fig:syndopu-drioct}(c) and \ref{fig:syndopu-drioct}(d)].
Because there are dense vasculatures at the choriocapillaris and choroid, the speckle pattern at these places might strongly depend on the duration of the acquisition.
As shown in [\figurenames~\ref{fig:syndopu-drioct}(a) and \ref{fig:syndopu-drioct}(c)], speckle appearance at this location is not the same.
It might be a possible source of the discrepancy.
Training with variable different averaging of OCT B-scans may stabilize the syn-DOPU appearances.

\begin{figure}
	\centering
	\includegraphics{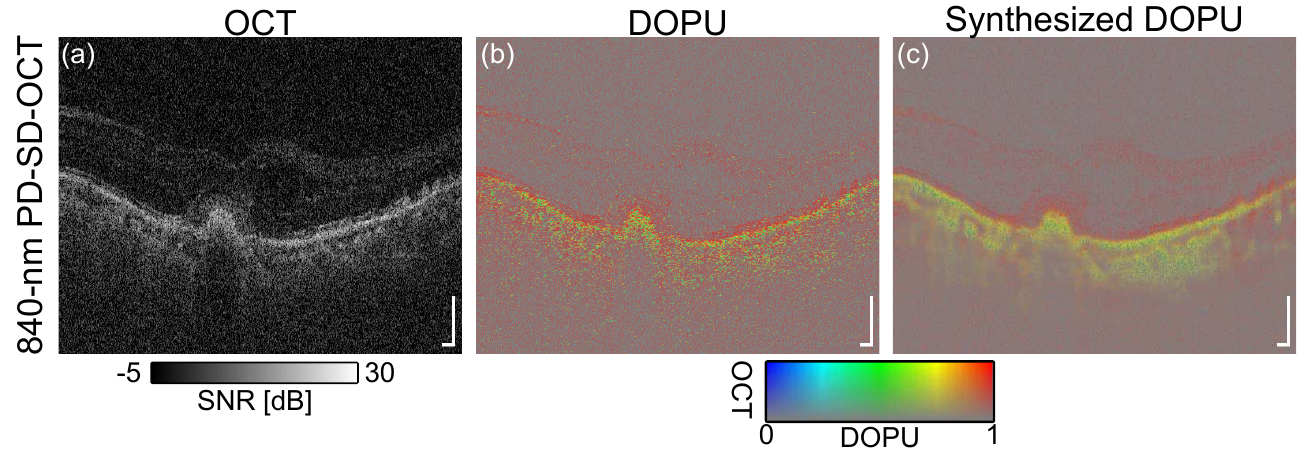}
	\caption{
		Syn-DOPU image of AMD patient obtained by 840-nm PD-SD-OCT.
		Scale bars indicate 200 \um.
	}\label{fig:syndopu-840nmPDSDOCT}
\end{figure}

\begin{figure}
	\centering
	\includegraphics{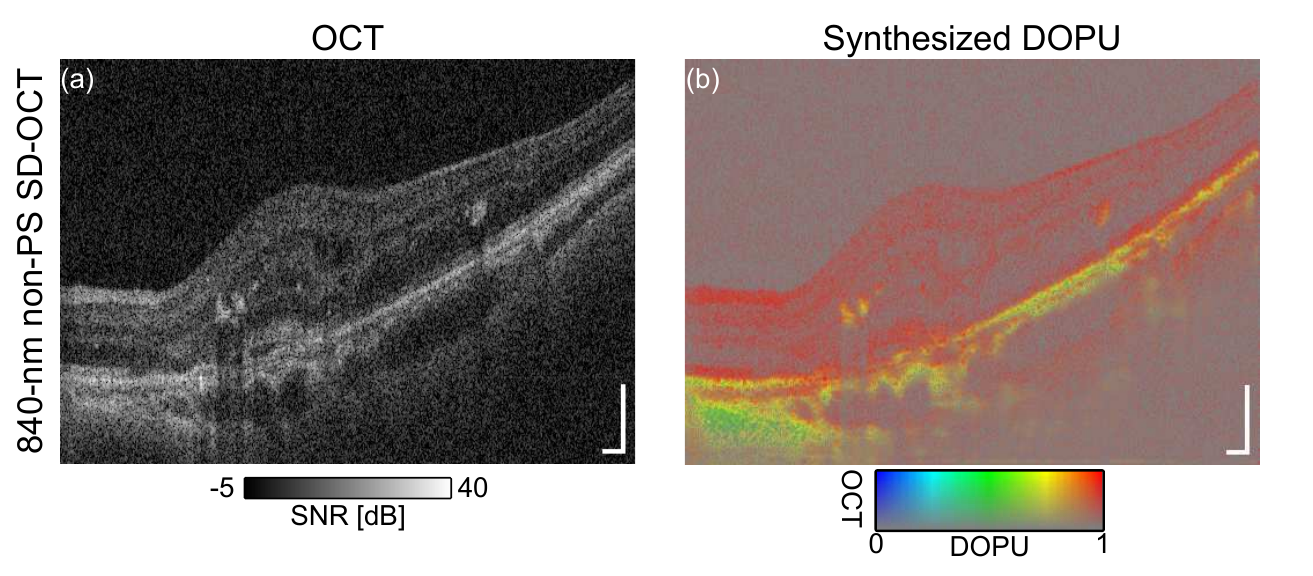}
	\caption{
		Syn-DOPU image of AMD patient obtained by 840-nm SD-OCT.
		Scale bars indicate 200 \um.
	}\label{fig:syndopu-840nmSDOCT}
\end{figure}

The trained model was also applied to OCTs with another wavelength range.
Some examples of Syn-DOPU images of AMD patients obtained from two 840-nm spectral-domain OCT (SD-OCT) setups are shown in \figurenames~\ref{fig:syndopu-840nmPDSDOCT} and \ref{fig:syndopu-840nmSDOCT}.
One is 840-nm PD-SD-OCT presented in Ref.~\cite{yamanari_fiber-based_2006} without input polarization state modulation and another is 840-nm non-PS SD-OCT presented in Ref.~\cite{makita_optical_2006}.
The spatial sampling properties of the images with 840-nm PD-SD-OCT are approximately 12 \um/A-line and 3 \um/pixel, and with 840-nm non-PS-SD-OCT are approx. 10 \um/A-line and 3 \um/pixel, respectively.
Both data were obtained without repetition of B-scan (\(N\) = 1), hence, both B-scan OCT intensity images [\figurenames~\ref{fig:syndopu-840nmPDSDOCT}(a) and \ref{fig:syndopu-840nmSDOCT}(a)] and DOPU image [\figurename~\ref{fig:syndopu-840nmPDSDOCT}(b)] are not averaged.
The DOPU image [\figurename~\ref{fig:syndopu-840nmPDSDOCT}(b)] contrasts the RPE, choroid, and exudative lesion.
However, it is noisy because a single B-scan with moderate SNR was used.
In contrast, the Syn-DOPU image [\figurename~\ref{fig:syndopu-840nmPDSDOCT}(c)] shows these structures smoothly.
The Syn-DOPU image of 840-nm non-PS SD-OCT [\figurename~\ref{fig:syndopu-840nmSDOCT}(b)] exhibits low DOPU values at choroidal stroma and some hyperreflective spots in the retina.
Also, the low DOPU values at the RPE except in some locations of the RPE where non-regular structures are shown in the OCT intensity [\figurename~\ref{fig:syndopu-840nmSDOCT}(a)].
These suggest that the proposed training approach is feasible to transfer other non-PS OCT devices.

\subsection{Related DOPU synthesis work}

Synthesizing PS-OCT images from a standard OCT has also been demonstrated by Sun \etal~\cite{sun_synthetic_2021}.
They trained a pix2pix GAN\cite{isola_image--image_2017} with the PS-OCT dataset and synthesized PS-OCT images, including DOPU and phase retardation, from OCT intensity images.
However, different samples and evaluation methods were used in that study.

Sun \etal~~\cite{sun_synthetic_2021} used human breast tissue, chicken skin tissue, and chicken muscle tissue as the samples.
Because these tissues are not melanin-rich, the tissue properties generating the DOPU probably differ from those of the retina.
In these samples, the randomization of the polarization state due to multiple scattering might be altered by interacting with birefringent tissues~\cite{wang_propagation_2002}, and the polarization state may be further randomized by randomly oriented birefringent tissues~\cite{jacques_imaging_2000}, i.e., collagen and muscle.
This also affects the OCT intensity image and hence provides features that may be used to synthesize the DOPU\@.

The similarity of synthesized and real images, as well as the image-wise classification of human breast tissues, were evaluated in Ref.~\cite{sun_synthetic_2021}.
In this study, we evaluated the syn-DOPU images based on the potential clinical biomarkers of the RPE abnormalities.
This is more directly related to the performance of the detailed diagnosis and investigation.
In any case, both studies show that synthesizing the DOPU from OCT intensity is feasible.
Therefore, synthesizing the DOPU is a promising way to extract the hidden features of tissues in OCT-intensity images.

\section{Conclusion}

An NN-based DOPU estimation algorithm was demonstrated.
OCT intensity images were input to a NN, and the corresponding DOPU images were synthesized.
The performance of the method was evaluated with an OCT dataset of healthy and pathological eyes.
Finally, the syn-DOPU images are similar to the DOPU images, and their ability to show the RPE abnormalities was in good agreement.
Synthesizing the DOPU from OCT intensity showed potential for use in clinical diagnosis.

\appendix
\section*{Appendix}

\section{Single-channel OCT emulation}\label{sec:1ch-emulation}

\begin{figure}
	\centering
	\includegraphics{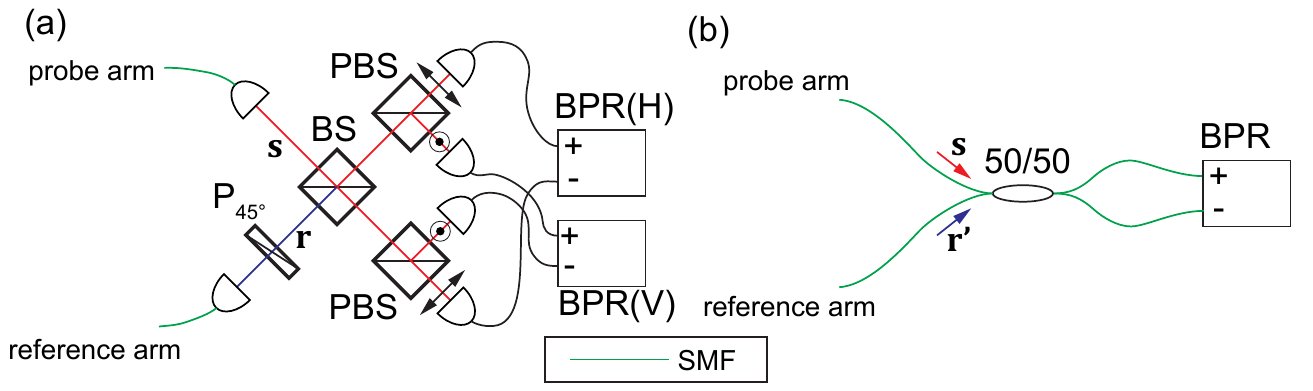}
	\caption{The schematic diagrams of interferometers and receivers of (a) PD-OCT and (b) the single-channel OCT.
	SMF: single-mode fiber, P\(_{45^\circ}\): 45\(^\circ\) linear polarizer, BS: beam splitter, PBS: polarizing beam splitter, BPR: balanced photoreceiver.}\label{fig:PDR-1chOCT}
\end{figure}

In the case of PD-OCT detection [\figurename~\ref{fig:PDR-1chOCT}(a)], the backscattered light from the probe arm with an unknown polarization state \(\mathbf{s} = [s_\mathrm{H}(x,k), s_\mathrm{V}(x,k)]^T\) and the reference light with 45\(^\circ\) linear polarization state \(\mathbf{r} = \frac{r(k)}{\sqrt{2}}[1,1]^T\) have interfered and H- and V-components are separately detected.
The polarization state of the reference light is clean-upped as 45\(^\circ\) by a polarizer to have equal powers on two receivers.
OCT signals of two channels are obtained from the interference term of backscattered and reference lights detected at two orthogonally-polarized channels.
\begin{align}\label{eq:pd-oct_signal}
	g_\mathrm{H}(x,z) = \mathcal{F}_k [s_\mathrm{H} (x, k) \frac{r^{*}(k)}{\sqrt{2}}](z), &&
	g_\mathrm{V}(x,z) = \mathcal{F}_k [s_\mathrm{V} (x, k) \frac{r^{*}(k)}{\sqrt{2}}](z),
\end{align}
where \(\mathcal{F}_k[\cdot](z)\) is the Fourier transform from \(k\)-domain to \(z\)-domain.

Assume that a conventional fiber-optic OCT with a single balanced receiver was used instead [\figurename~\ref{fig:PDR-1chOCT}(b)], the detected single-channel spectral intensity can be described with a reference light of an arbitrary state \(\mathbf{r'} = [r'_\mathrm{H}(k), r'_\mathrm{V}(k)]^T\) as follows:
\begin{equation}\label{eq:1ch_spectral_intensity}
	\begin{split}
		S_\mathrm{1ch}(x,k) =& | \mathbf{s}(x,k) + \mathbf{r^{'}}(k) |^2\\
		=& \left|
		\begin{bmatrix}
			s_\mathrm{H} (x, k) + r'_\mathrm{H}(k)\\
			s_\mathrm{V} (x, k) + r'_\mathrm{V}(k)
		\end{bmatrix}
		\right|^2\\
		=& |s_\mathrm{H} (x, k) + r'_\mathrm{H}(k)|^2 + |s_\mathrm{V} (x, k) + r'_\mathrm{V}(k)|^2\\
		=& |s_\mathrm{H} (x, k)|^2 + |r'_\mathrm{H}(k)|^2 + |s_\mathrm{V} (x, k)|^2 + |r'_\mathrm{V}(k)|^2\\
		 & + s_\mathrm{H} (x, k) r_\mathrm{H}^{'*}(k) + c.c. + s_\mathrm{V} (x, k) r_\mathrm{V}^{'*}(k) + c.c.
	\end{split},
\end{equation}
where \(c.c.\) is the complex conjugate of the interference terms.
OCT signal is the Fourier transform of the interference terms.
Thus, the complex OCT signal in the case of single-channel detection is
\begin{equation}\label{eq:def_1ch_OCT_signal}
	\begin{split}
		g_\mathrm{1ch}(x,z) =& \mathcal{F}_k [s_\mathrm{H} (x, k) r_\mathrm{H}^{'*}(k) + s_\mathrm{V} (x, k) r_\mathrm{V}^{'*}(k)](z)\\
					 =& g'_\mathrm{H}(x,z) + g'_\mathrm{V}(x,z)
	\end{split}.
\end{equation}
The relationship between the arbitrary polarization state \(\mathbf{r'}\) and \(\mathbf{r}\) can be described as:
\begin{equation}
	\mathbf{r'} = \mathbf{U}\cdot\mathbf{r},
\end{equation}
where the matrix \(\mathbf{U}\) represents an arbitrary transformation of a polarization state without changing the total light intensity.
Hence, \(\mathbf{U}\) is a \(2\times2\) unitary matrix\cite{whitney_pauli-algebraic_1971} and can be expressed as:
\begin{equation}
	\mathbf{U} = \begin{bmatrix}
		\alpha & \beta\\
		- e^{i \theta}\beta^* & e^{i \theta} \alpha^*
	\end{bmatrix},
\end{equation}
where \(|\alpha|^2 + |\beta|^2\) = 1.
Since \(\operatorname{det}[\mathbf{U}] = e^{i \theta}\), its magnitude is unity \(|\operatorname{det}[\mathbf{U}]| = 1\).
The polarization components of the reference light are then
\begin{align}
	r'_\mathrm{H}(k) = (\alpha + \beta) \frac{r(k)}{\sqrt{2}}, && r'_\mathrm{V}(k) = (\alpha^* - \beta^*)e^{i\theta} \frac{r(k)}{\sqrt{2}}.	
\end{align}
The Eq.~(\ref{eq:def_1ch_OCT_signal}) can be rewritten as:
\begin{equation}\label{eq:single-channel-OCT-emulation}
	\begin{split}
		g_\mathrm{1ch}(x,z) =& A g_\mathrm{H}(x,z)
		+ B g_\mathrm{V}(x,z)
	\end{split},
\end{equation}
where \(A = \alpha^* + \beta^*\) and \(B = (\alpha - \beta)e^{-i\theta}\)
Hence, the single-channel OCT signal \(g_\mathrm{1ch}\) can be emulated by a linear combination of two PD-OCT signals \(g_\mathrm{H}\) and \(g_\mathrm{V}\).
Eq.~(\ref{eq:def_OCT_intensity}) is corresponding to the single-channel conventional OCT intensity when \(\alpha = 1\), \(\beta = 0\), and \(\theta = \Delta \phi_\mathrm{ch}\).

\section{SNR penalizing based on OCT intensity}\label{sec:snr-penalty}

To emulate an OCT signal with higher levels of additive noise, the straightforward approach is to add a complex Gaussian random value to the complex OCT signal.
However, we need to address raw complex OCT data.
Hence, a method of low-SNR OCT intensity emulation using OCT intensity was implemented.

The SNR-reduced OCT intensity by adding random zero-mean Gaussian variables to complex OCT signal is described as
\begin{equation}\label{eq:snr-penalty-complex}
	\begin{split}
		|g'|^2 &= |g_\mathrm{re} + n_\mathrm{re} + i (g_\mathrm{im} +n_\mathrm{im})|^2 \\
			   &= (g_\mathrm{re} +n_\mathrm{re}  )^2+(g_\mathrm{im} +n_\mathrm{im})^2 \\
			   &= |g|^2 + 2 g_\mathrm{re} n_\mathrm{re} + 2 g_\mathrm{im} n_\mathrm{im} + (n_\mathrm{re}^2 + n_\mathrm{im}^2)
	\end{split},
\end{equation}
where \(g'\) and \(g=g_\mathrm{re} + i g_\mathrm{im}\) are the emulated and measured complex OCT signals, and \(n_\mathrm{re}\) and \(n_\mathrm{im}\) are the additional additive noises for the real and imaginary parts.

\begin{figure}
	\centering
	\includegraphics{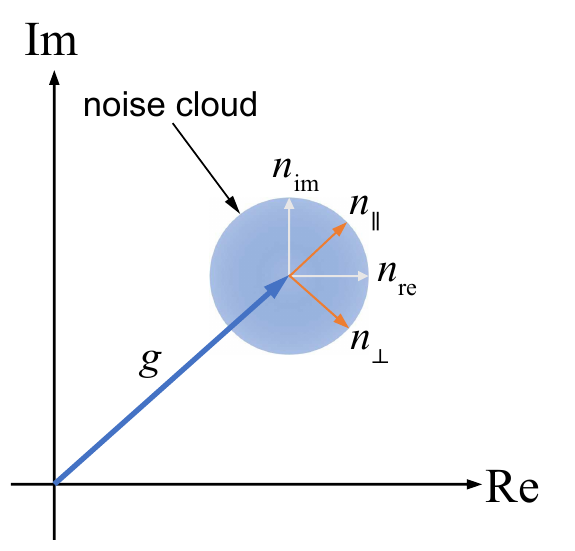}
	\caption{Relationship between the measured complex OCT signal \(g\) and additional additive noises \(n\).}\label{fig:snr-penalty}
\end{figure}

If we know only the OCT intensity, we have only the magnitude of \(g\) and simulated noises \(n_\mathrm{re}\) and \(n_\mathrm{im}\).
The challenge is to calculate the 2nd and 3rd terms of Eq.~(\ref{eq:snr-penalty-complex}) using OCT intensity, namely, to assume the splitting ratio of OCT intensity into real and imaginary parts.
Fortunately, we do not need to know about the ratio.
If we assume complex circular Gaussian noise, i.e., \(n_\mathrm{re}\) and \(n_\mathrm{im}\) are independent Gaussian random variables, the selection of the coordinates of the complex plane does not change the statistics of Eq.~(\ref{eq:snr-penalty-complex}).
Hence, we can choose the simplest case, one axis along the measured signal phasor \(g\) and another perpendicular to that [\figurename{}~\ref{fig:snr-penalty}].
Using the newly selected coordinates, we can emulate low-SNR OCT intensity with additional additive noises in the complex signal as
\begin{equation}\label{eq:snr-penalty-intensity}
	I' = I + 2 \sqrt{I} \cdot n_\parallel + n_\parallel^2 + n_\perp^2,
\end{equation}
where \(I=|g|^2\) is the measured OCT intensity signal.
\(n_\parallel\) and \(n_\perp\) are random Gaussian variables parallel and perpendicular to measured signal phasor \(g\).
By generating two independent random zero-mean Gaussian variables and substituting them into \(n_\parallel\) and \(n_\perp\) of Eq.~(\ref{eq:snr-penalty-intensity}), an SNR-reduced OCT intensity signal can be simulated using only the magnitude of the measured OCT signal \(\sqrt{I} = |g|\).

\begin{backmatter}
	\bmsection{Funding}
	Core Research for Evolutional Science and Technology (JPMJCR2105);
	Japan Society for the Promotion of Science (JSPS) KAKENHI (18H01893, 18K09460, 21K09684, 21H01836, 22K04962).

	\bmsection{Acknowledgments}
	We acknowledge Takuya Iwasaki, Tokyo Medical University Ibaraki Medical Center, for providing grading help, Tatsuo Yamaguchi, Topcon Corporation, for technical support and discussions, and Kensuke Oikawa, for the preparation of codes.

	\bmsection{Disclosures}

	\noindent
	SM, YY:\@ Topcon (F), Yokogawa Electric (F), Nikon (F), Sky Technology (F), Kao (F).
	MM:\@ Santen (F).
	SA, TM:\@ Topcon (E).

	\bmsection{Data availability} Data underlying the results presented in this paper are not publicly available at this time but may be obtained from the authors upon reasonable request.

\end{backmatter}

\end{document}